\documentstyle[prl,aps,epsfig,multicol]{revtex}

\newcommand{\eqref}[1]{(\ref{#1})}

\begin{document}


\draft

\title{Maternal effects in molecular evolution}

\author{Claus O. Wilke}

\address{Digital Life Laboratory, Mail Code 136-93,\\Caltech, Pasadena, CA 91125\\wilke@caltech.edu}

\date{Printed: \today}

\maketitle

\begin{abstract}
  We introduce a model of molecular evolution in which the fitness of an
  individual depends both on its own and on the parent's genotype. The model
  can be solved by means of a nonlinear mapping onto the standard quasispecies
  model. The dependency on the parental genotypes cancels from the mean
  fitness, but not from the individual sequence concentrations. For finite
  populations, the position of the error threshold is very sensitive to the
  influence from parent genotypes. In addition to biological applications, our
  model is important for understanding the dynamics of self-replicating
  computer programs.
\end{abstract}


\begin{multicols}{2}

  Simple models of asexual evolution, such as the quasispecies model,
  typically assume that the fitness of an organism is a function of only its
  genotype and the environment. This allows for the analysis of evolution in
  static~\cite{EigenSchuster79,Eigenetal89,vanNimwegenetal99a,vanNimwegenCrutchfield2000,PBennetShapiro1994,RattrayShapiro96,RogersPBennet2000}
  or variable~\cite{NillsonSnoad2000,WilkeRonnewinkel2001,Wilkeetal2001a}
  environments, many to one mappings from genotype to phenotype
  (neutrality)~\cite{Huynenetal96,Reidysetal2001,vanNimwegenetal99b}, and
  phenotypic plasticity~\cite{AncelFontana2000}. These models disregard the
  potential influence of the mother (or parent, in general) on the organism's
  phenotype. This influence comes about because in addition to the genetic
  material, a wealth of proteins and other substances is transferred from
  mother to child. In the context of sexual reproduction, the influence of the
  mother on a child's phenotype is usually called a maternal effect. A classic
  example is that of the snail \emph{Partula suturalis}~\cite{Dawkins82}, for
  which the directionality in the coiling of the shells is determined by the
  genotype of the mother of an organism, rather than the organism's own
  genotype.  Maternal effects are not exclusive to sexually reproducing
  organisms, however, they can be observed in simple asexual organisms as
  well. In bacteria, for example, the fitness of a strain in a given
  environment may depend on the environment that was experienced by the
  strain's immediate ancestors~\cite{Leroietal94,Lenskietal94}.

Here, our objective is to define and study a model of the evolution of asexual
organisms that takes such maternal effects into account. We assume that the
fitness of an organism is given by the product of two quantities $a$ and $b$,
where $a$ depends solely on the genotype of the mother of the organism, and
$b$ depends solely on the organism's own genotype. Since we need to keep track
of the abundances of all possible mother/child combinations, we
need $n^2$ concentration variables if we distinguish $n$ different genotypes
in our model. In the following,  we denote by $x_{ij}$ the concentration of
organisms of genotype $j$ descended from genotype $i$, and by $q_{ji}$
the probability that a genotype $i$ mutates into genotype $j$. The time
evolution of the $x_{ij}$ is then, in analogy to the quasispecies model,
\begin{equation}\label{maternal-eff}
  \dot x_{ij}(t) = \sum_k a_kb_i q_{ji}x_{ki}(t) - f(t)x_{ij}(t)\,,
\end{equation}
with $f(t)=\sum_{i,j}a_ib_jx_{ij}(t)$. The function $f(t)$ gives the average
fitness of the population at time $t$. In principle, this model can be solved
by diagonalizing a $n^2\times n^2$ matrix. However, since even for relatively
short genetic sequences the number of possible genotypes $n$ is enormous, this
direct method is very cumbersome. Fortunately, a simple transformation exists
that reduces the above problem to one in which the diagonalization of a
$n\times n$ matrix is sufficient. Namely, if we introduce variables $x_i$ such
that
\begin{equation}\label{trans}
  x_{ij} = b_i q_{ji} x_i\Big/\Big(\sum_{k} b_k x_k\Big)\,,
\end{equation}
then, after inserting Eq.~\eqref{trans} into
Eq.~\eqref{maternal-eff}, we obtain in the steady state ($\dot x_{ij}=0$)
\begin{equation}\label{quasispecies}
  \tilde f x_i = \sum_j a_jb_jq_{ij} x_j
\end{equation}
with $\tilde f=\sum_j a_jb_j x_j$. The inverse transformation, which converts
Eq.~\eqref{quasispecies} back into the right-hand-side of
Eq.~\eqref{maternal-eff}, can be achieved with
\begin{equation}\label{back-trans}
  x_i = \Big(\sum_j a_j x_{ji}\Big)\Big/\Big(\sum_{j,k} a_j x_{jk}\Big)\,.
\end{equation}
Therefore, Eq.~\eqref{quasispecies} is fully equivalent to the steady state of
Eq.~\eqref{maternal-eff}. This leads to an interesting conclusion. Note that
Eq.~\eqref{quasispecies} is simply the steady state equation of the
quasispecies model if we assume that genotypes $j$ replicate with replication
rate $c_j=a_jb_j$ and mutate into genotypes $i$ with $q_{ij}$, while $x_i$
gives the relative concentration of genotype $i$. Consequently, the model with
maternal effects is mathematically equivalent to the standard quasispecies
model. Moreover, with the aid of Eq.~\eqref{trans} and Eq.~\eqref{back-trans},
it can be shown that $f(t\rightarrow\infty)=\tilde f$. Therefore,
the average fitness in both models is the same; the maternal effects
drop out of the expression for the average fitness.

While the average fitness depends only on the values of $c_i$, the individual
sequence concentrations actually depend on the details of the maternal
effects.  In particular, the total amount of sequences of a given genotype $i$
is not identical to the corresponding value $x_i$ in the standard quasispecies
model ($\sum_j x_{ji} \neq x_i$ in general).  From Eq.~\eqref{trans}, we see
that for every given mutation matrix $q_{ij}$, we can suppress any sequence
concentration to as small a level as we please, by reducing the corresponding
$b_i$ and holding the other $b_j$ constant ($\lim_{b_i\rightarrow0} x_{ij} =
0$). Enhancement of sequence concentrations is also possible, although there
exists an upper bound that cannot be exceeded. The upper bound is given by the
matrix element $q_{ji}$ ($\lim_{b_i\rightarrow\infty} x_{ij} = q_{ji}$). Its
existence is easy to understand: by changing $b_i$, we can only jointly
manipulate the concentrations of \emph{all} sequences descended from genotype
$i$. The ratio between different genotypes $j$ descended from $i$ is always
fixed, and it is determined by the matrix elements $q_{ji}$. At most, the sum
over all descendants from all genotypes $i$ in the population can be one,
$\sum_j x_{ij} = 1$, which implies $x_{ij}=q_{ji}$.

We will now classify the different types of maternal effects.  If all $a_i=1$,
such that $b_i=c_i$, no maternal effects are present, and we obtain the normal
sequence concentrations from the quasispecies model. We will refer to this
situation as the \emph{neutral} case. In order to classify all non-neutral
situations, we compare concentrations of those sequences that are true copies
of their parents (this is the only reasonable way, given that \emph{all}
genotypes descended from the same ancestor $i$ scale identically with $b_i$).
If the concentration $x_{ii}$ of a sequence with large $c_i$ is
\emph{reduced}, while the concentration $x_{jj}$ of some other sequence with
smaller $c_j$ is \emph{enhanced}, we will speak of \emph{positive} maternal
effects. Likewise, if the sequence concentration of a faster replicating
sequence is \emph{enhanced}, at the expense of some slower replicating
sequence, we will speak of \emph{negative} maternal effects. In short,
positive maternal effects promote slower replicators, and negative maternal
effects promote faster replicators. We refer to the above classification as
the \emph{direction} of the maternal effects. Likewise, the \emph{strengh} of
the maternal effects indicates the degree to which a system deviates from the
neutral case (\emph{weak} maternal effects show only a small deviation,
\emph{strong} maternal effects show a large deviation from the neutral case).

With Eq.~\eqref{trans} we can solve the model, as long
as there exists an analytical solution for the corresponding quasispecies
landscape. This means that solutions for multiplicative
landscapes~\cite{Rumschitzki87,WoodcockHiggs96}, the single peak
landscape~\cite{Galluccio97}, and certain spin-glass
landscapes~\cite{Leuthaeusser87,Tarazona92,Franzetal93} are readily available.
In the following, we discuss the well-known example of the sharp single-peak
landscape~\cite{SwetinaSchuster82}. We assume that the genetic sequences are
binary of length $\ell$.  Moreover, we assume a uniform copy fidelity $q$ per
digit. The sequence $000\dots0$ may replicate (in the absence of mutations)
with rate $c_0=a_0b_0$. We will refer to this sequence as the master sequence.
Let all other sequences replicate with $c_1=a_1b_1 \ll c_0$.  If $\ell$ is
large, we may neglect back-mutations onto the master sequence, in which case it
is sufficient to keep track of the total concentration of all sequences off
the peak in a single variable, $x_1$. The mutation matrix $q_{ij}$ is then a
$2\times2$ matrix with the elements $q_{00}=q^\ell$, $q_{10}=1-q^\ell$,
$q_{01}=q_{10}/(2^\ell-1)$, and $q_{11}=1-q_{01}$ (the elements $q_{01}$ and
$q_{01}$ are approximated). In the standard quasispecies model, the
equilibrium concentration of the master sequence $x_0$ is given by
\begin{equation}\label{master-seq-conc}
  x_0 = (c_0 q_{00} - c_1)/(c_0-c_1)\,,
\end{equation}
and $x_1$ likewise as $x_1=1-x_0$. The average fitness follows as
\begin{equation}
  \label{ave-fitness}
  f = \left\{\begin{array}{r@{\quad}l} c_0 q_{00} & \mbox{for $x_0\geq 0$,}\\
        c_1 & \mbox{else.} \end{array}\right.
\end{equation}
Now, for the sequence concentrations with maternal effects, we obtain from
Eq.~\eqref{master-seq-conc} in conjunction with Eq.~\eqref{trans}
\begin{eqnarray}\label{pred1}
  x_{0i} &=& b_0q_{i0} (c_0q_{00}-c_1)/\Lambda\,,\\
  x_{1i} &=& b_1q_{i1} (c_0-c_0q_{00})/\Lambda\,,\label{pred2}
\end{eqnarray}
with $\Lambda=(b_0-b_1)c_0q_{00} +b_1c_0-b_0c_1$ and $i=0,1$.

\begin{figure}
\widetext
\centerline{
\epsfig{file={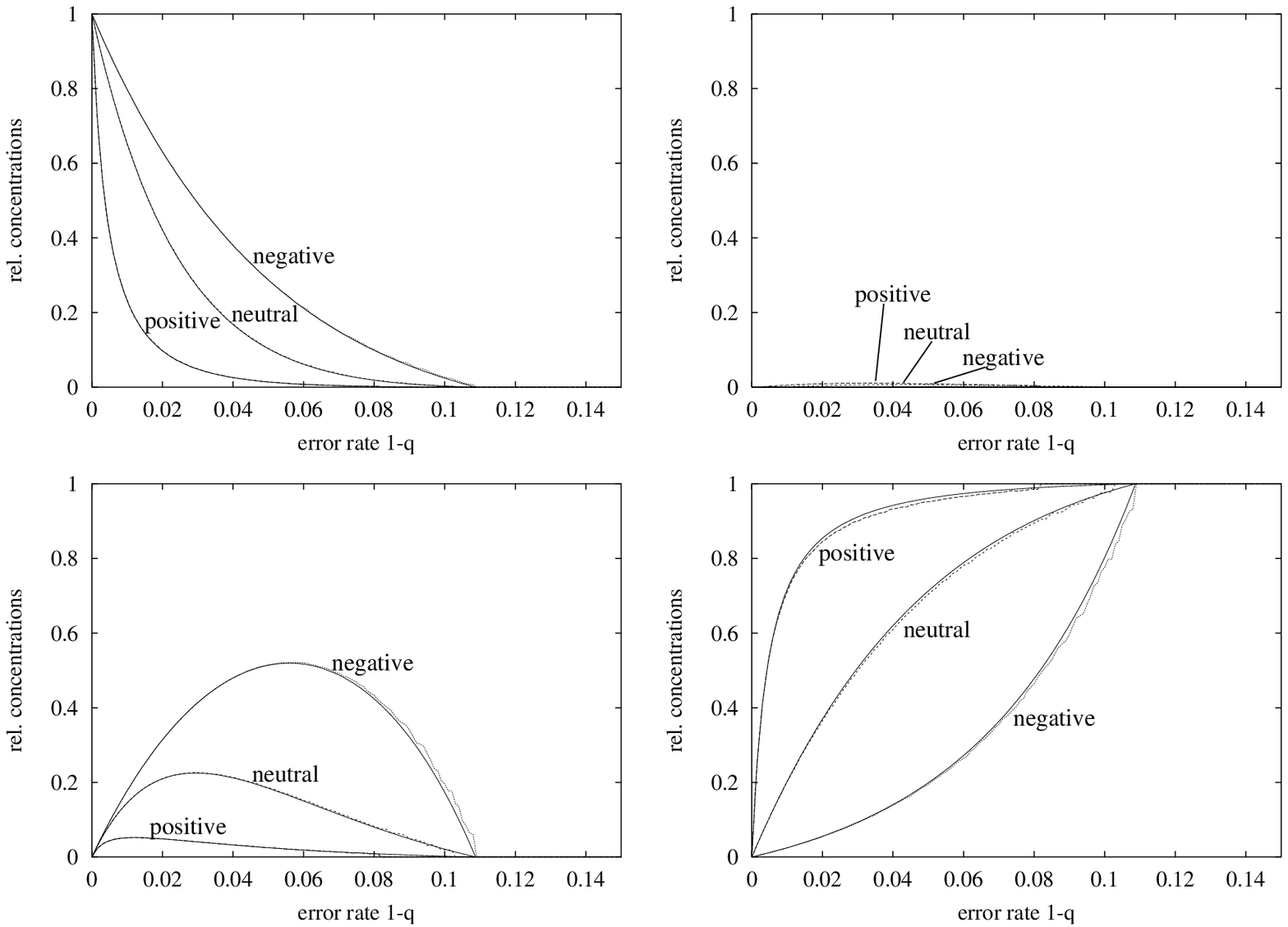}, width=.75\columnwidth}
}
\caption{\label{fig:rel_concs}Relative sequence concentrations vs.\ error rate
  $1-q$. From left to right, and from top to bottom, we display $x_{00}$,
  $x_{10}$, $x_{01}$, $x_{11}$. Solid lines are the analytical predictions
  Eqs. \eqref{pred1},~\eqref{pred2}, dashed lines stem from simulations with
  $N=10000$ sequences of length $l=20$. The parameters of the fitness
  landscapes were $c_0=10$ and $c_1=1$, with $b_0=0.1$, $b_1=1$ (positive);
  $b_0=1$, $b_1=1$ (neutral); $b_0=1$, $b_1=0.1$ (negative).}
\end{figure}

Figure~\ref{fig:rel_concs} displays the four sequence concentrations $x_{00}$,
$x_{01}$, $x_{10}$, $x_{11}$ of the above defined landscape, for positive,
negative, and neutral maternal effects. We see that indeed the maternal
effects result in a significant shift in the sequence concentrations.  The
concentration $x_{11}$ (shown in the lower right of Fig.~\ref{fig:rel_concs}),
e.g., encompasses almost the complete population for positive maternal effects
at an error rate of about 0.04, while it constitutes less than 20\% in the
case of the negative maternal effects for the same error rate.

\begin{figure}
\narrowtext
\centerline{
\epsfig{file={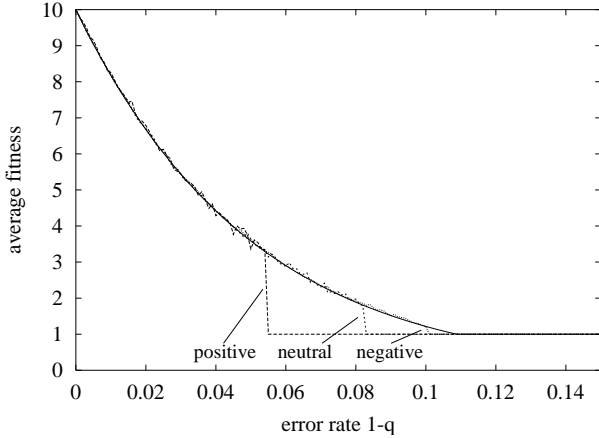}, width=\columnwidth}
}
\caption{\label{fig:ave_fitness}Average fitness vs.\ error rate $1-q$. The
  solid line represents Eq.~\eqref{ave-fitness}, and the dashed lines stem
  from simulations with $N=1000$ sequences of length $l=20$. The fitness
  landscapes were identical to Fig.~\ref{fig:rel_concs}.  }
\end{figure}

The potential shift in the individual sequence concentrations has important
implications for finite populations. When the concentration of a sequence (as
predicted for an infinite population) approaches the inverse of the population
size, that sequence will most certanily be lost through sampling
fluctuations. In the case of the master sequence, this effect is responsible
for the shift of the error catastrophe towards smaller error rates for finite
populations in the ordinary quasispecies
model~\cite{NowakSchuster89,Wieheetal95,AlvesFontanari98,CamposFontanari99}.
Now, since the concentration of the master sequence can be made arbitrarily
small with suitable maternal effects, it follows that the error threshold can
be shifted. This effect is illustrated in Fig.~\ref{fig:ave_fitness} for a
population size of $N=1000$, for which the error transition in the normal
quasispecies model (as represented by the 'neutral' case) is already
significantly shifted. Positive maternal effects increase this shift by a fair
amount, while negative maternal effects can almost completely counteract the
finite population effect, and move the error transition very close to the
infinite population limit. Besides the shift in the error transition,
Fig.~\ref{fig:ave_fitness} shows that the average fitness is indeed unaffected
by strength and/or direction of the maternal effects, as all three curves lie
exactly on the infinite population result for error rates below the respective
error transitions.

\begin{figure}
\narrowtext
\centerline{
\epsfig{file={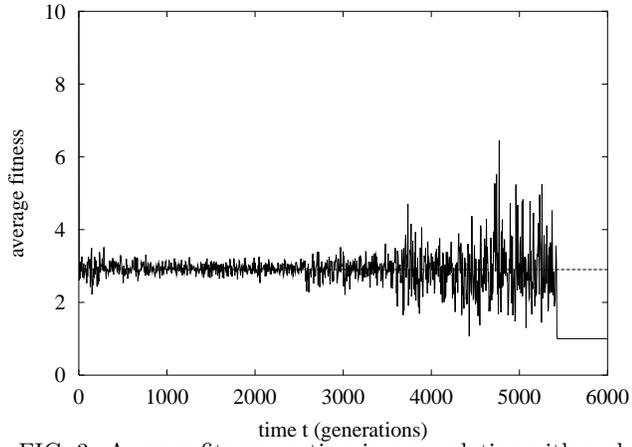}, width=\columnwidth}
}
\caption{\label{fig:evo}Average fitness vs.\ time in a population with
  evolving maternal effects. The dashed line indicates the infinite population
  result [Eq.~\eqref{ave-fitness}]. The population consisted of $N=1000$
  sequencees of $l=20$, the error rate was $1-q=0.06$, and the landscape was
  defined by $c_0=10$, $c_1=1$. The scale parameter $\alpha$
  was set to $\alpha=10$.  }
\end{figure}

We have seen above that the mean fitness in the population is not influenced
by the existence of maternal effects. Since selection acts only on the average
fitness~\cite{SchusterSwetina88,Wilkeetal2001b}, it follows that the maternal
effects cannot be under selective pressure. In order to verify this, we have
performed simulations in which strength and direction of the maternal effects
were allowed to evolve. To each sequence in the population, we added an
inheritable variable $z$. On reproduction, the offspring received a value
$z'=z+dz$, where $dz$ was a normally distributed random variable. For the
master sequence, $z$ was then mapped into $a_0$ and $b_0$ via $a_0 =
(\alpha+z)/\alpha$ for $z>0$, $a_0=\alpha/(\alpha+z)$ for $z\leq0$, and $b_0 =
1/a_0$, with $\alpha$ defining the scale between $z$ and $a_0$, $b_0$. For
$a_1$ and $b_1$, the value of $z$ was ignored.  Figure~\ref{fig:evo} shows a
typical simulation run in such a system. We chose $N=1000$ and $1-q=0.06$, so
that the population was below the error threshold in the absence of maternal
effects, and we initialized all sequences in the population to $z=0$. Over the
course of a simulation, the $z$ values drift randomly, which can be seen in
increasing and diminishing fluctuations about the average fitness.  When the
average $z$ drifts below zero, the fluctuations decrease, because $z<0$
corresponds to negative maternal effects, which shift the population away from
the error threshold.  When the average $z$ drifts above zero, on the other
hand, the fluctuations increase. If there is no upper limit to $z$, the
fluctuations will eventually grow so large that the population is pushed over
the error threshold. In Fig.~\ref{fig:evo}, this happend around generation
5400.

The model we have introduced in this paper oversimplifies the situation for
bacteria, where substances can remain in the cytoplasm for several
generations, such that not only the parent, but also the grand- and the
grand-grand-parent etc.\ have an influence on the phenotype of an individual.
However, it is an exact description of the dynamics of the digital organisms
(self-replicating and evolving computer programs) of the Avida system, which
has been used extensively in experimental evolution
research~\cite{Wilkeetal2001b,Adami98,Lenskietal99,Adamieetal2000,WagenaarAdami2000,OfriaAdami2001}.
The replication rate of these digital organisms is the ratio between the
number of instructions per unit time that they can execute [the speed of their
central processing unit (CPU)] and the number of instructions they have to
execute in order to produce a new offspring (length of the gestation cycle).
The CPU speed depends on the number and type of logical operations that these
organisms perform in addition to their replicatory activity (the more logical
operations an organism performs, the faster its CPU will operate).
Since the logical operations an organism can perform are only known \emph{a
  posteriori}, these organisms obtain their initial CPU speed from their
parent organism. The CPU speed corresponds thus to the parameter $a$ of the
present work, and the length of the gestation cycle to the inverse of the
parameter $b$. Therefore, we have shown in that a quasispecies
description of the digital organisms is indeed justified, as was proposed
in~\cite{Wilkeetal2001b}. Also, our model might lead to a detailed
quantitative description of the dynamics of digital organisms in future work.

This work was supported by the National Science Foundation under Contract No.
DEB-9981397. The author thanks Chris Adami for carefully reading this
manuscript.


\end{multicols}

\newpage

\end{document}